\documentclass{aa}  
\usepackage{graphicx}
\usepackage{txfonts}
\usepackage{natbib}
\bibpunct{(}{)}{;}{a}{}{,} % to follow the A&A style

\begin{document}

   \title{Optical microflares in LS I +61 303 and the search for their multiwavelength counterpart}
   \titlerunning{Microflares in LS I +61 303}
    \authorrunning{Mestre et al.}
    %\subtitle{} 

   \author{Enrique  Mestre\inst{1}, Estrella S{\'a}nchez-Ayaso\inst{2}, Pedro L. Luque-Escamilla\inst{3} Josep Mart{\'i}\inst{1}, 
   Josep M. Paredes\inst{4}, Daniel del Ser\inst{4,5}, \and Jorge N\'u\~nez\inst{4,5}
          }
          
   \institute{
   Departamento de F{\'i}sica (EPSJ), Universidad de Ja{\'e}n, Campus Las Lagunillas s/n, A3, E-23071 Ja{\'e}n, Spain\\
             \email{emestre@ujaen.es}
   \and
            Departamento de Ciencias Integradas, Centro de Estudios Avanzados en F{\'i}sica, Matem{\'a}tica y Computaci{\'o}n, Universidad de Huelva, E-21071, Huelva, Spain
%\email{estrella.sanchez@dci.uhu.es}
\and
   Departamento de Ingenier{\'i}a Mec{\'a}nica y Minera (EPSJ), Universidad de Ja{\'e}n, Campus Las Lagunillas s/n, A3, E-23071 Ja{\'e}n, Spain
 %             \email{peter@ujaen.es}
%         \and
%             Departamento de F{\'i}sica (EPSJ), Universidad de Ja{\'e}n, Campus Las Lagunillas s/n, A3, E-23071 Ja{\'e}n, Spain\\
 %            \email{jmarti@ujaen.es}
 \and
 Departament de F\'{\i}sica Qu\`antica i Astrof\'{\i}sica, Institut de Ci\`encies del Cosmos (ICCUB), Universitat de Barcelona, IEEC-UB, Mart\'{\i} i Franqu\`es 1, E-08028 Barcelona, Spain
\and
Observatori Fabra, Reial Acad\`emia de Ci\`encies i Arts de Barcelona, Rambla dels Estudis, 115, E-08002 Barcelona, Spain
}

   \date{Received 12/02/2022; accepted 29/04/2022}

% \abstract{}{}{}{}{} 
% 5 {} token are mandatory
 
  \abstract
  {Stellar sources of gamma rays are one of the front lines in modern astrophysics whose understanding can benefit from observational tools not originally
   designed for their study.}
   {We take advantage of the high precision photometric capabilities of present-day space facilities to obtain a new perspective
   on the optical behavior of the X-ray and gamma-ray binary LS I +61 303.  Previously unknown phenomena whose effects manifest with amplitudes below 0.01 magnitude
   can now be  clearly observed and studied. }
   {Our work is mainly based on the analysis of  optical and gamma-ray archival data and uses the tools recommended by the different collaborations
   that provide these valuable observational  resources (in particular, the {\it TESS} and {\it Fermi} orbiting observatories).  In addition, complementary ground-based optical spectroscopy has also been conducted.}
   {We report the discovery of small-amplitude optical flares on timescales of a day in the LS I +61 303 light curve. Different alternative scenarios to explain
   their  
   origin are tentatively proposed.}
   {}

   \keywords{stars: flare -- gamma rays: stars -- X-rays: binaries --  stars: emission line, Be -- stars: individual: LS I +61 303
    }

   \maketitle
%
%-------------------------------------------------------------------
 
\section{Introduction}

LS I +61 303 is a high mass X-ray binary that consists of a B0 Ve star \citep{2005MNRAS.360.1105C} and a compact object whose nature 
has remained unknown for decades. 
Very recently, the detection of radio pulsations provides evidence in favor of a rotating neutron star
and against the alternative black hole option \citep{2022NatAs.tmp...71W}. Its distance was first estimated as
$\sim 2$ kpc \citep{1991AJ....101.2126F}. 
The binary system orbital parameters best indicate a highly eccentric orbit ($e = [0.5-0.72]$) with periastron at phase $[0.23-0.3]$ \citep{1981PASP...93..486H,2005MNRAS.360.1105C, 2007ApJ...656..437G}.
% where phase zero dates back to the first detection at radio \citep{1978Natur.272..704G,2002ApJ...575..427G}. 
Be stars present slow equatorial outflows in the form of a thin disk \citep{1988A&A...198..200W}. In the case of LS I +61 303, the companion star mass has been estimated as 10 to 15 $\rm{M}_{\odot}$ for an orbital inclination $i < 60\degr{}$ \citep{2005MNRAS.360.1105C}. %with radius $\approx 7\ \rm{R}_{\odot}$  

The high radio variability of LS I +61 303 was first reported in 1978, during a Galactic plane survey for highly variable radio sources (\citealt{1978Natur.272..704G}). 
Soon after, a $\sim 26.5$ day periodic modulation of its radio emission, corresponding to the binary system orbital period, 
was noticed \citep{1982ApJ...255..210T, 2002ApJ...575..427G}. The commonly used zero phase epoch  dates back to the first detection in the radio \citep{1978Natur.272..704G}.
Super-orbital radio variability was also found with a $\sim 4$ year 
modulation of the radio outburst peak \citep{1987PhDT.......113P,   1995A&A...298..151M,2002ApJ...575..427G}. In contrast, small-amplitude radio variability has been reported down to timescales of $\approx 1.4$ hours during the flux decay soon after the expected periodic radio outburst \citep{1997A&A...328..283P}. 
The existence of precessing collimated radio jets has been claimed from high-resolution ($20-50$ milliarcseconds) radio interferometric observations \citep{2004A&A...414L...1M,2001A&A...376..217M}, 
which would support a microquasar scenario. Multiple models (both leptonic and hadronic) have been proposed within this context to account for the broadband emission of LS I +61 303 \citep{2006A&A...459L..25B,Gupta2006ATL,dermer2006gamma,bednarek2006inverse,Bednarek:2006nc,Romero2005HadronicHG,torres2007ls}.
Other competing scenarios proposed to explain the emission from the source are the {``colliding winds''} models, in which the compact object is a young pulsar losing energy via relativistic wind 
\citep{1981MNRAS.194P...1M,2006A&A...456..801D,Romero2007AccretionVC,2010MNRAS.403.1873Z}. A shock forms in the boundary between the pulsar wind and the companion stellar wind, where synchrotron emitting leptons can be accelerated to later interact via inverse Compton with the Be companion photon field.
%\textbf{creo que debemos aniadir a esta frase algo que diga que existen otras interpretaciones como pulsar+Be. Lo digo por si el referee es afecto a esta ultima version...}

 %--------------------------------------------------------------------

   \begin{figure*}[t!]
      \resizebox{\hsize}{!}
        {\includegraphics[width=8.0cm]{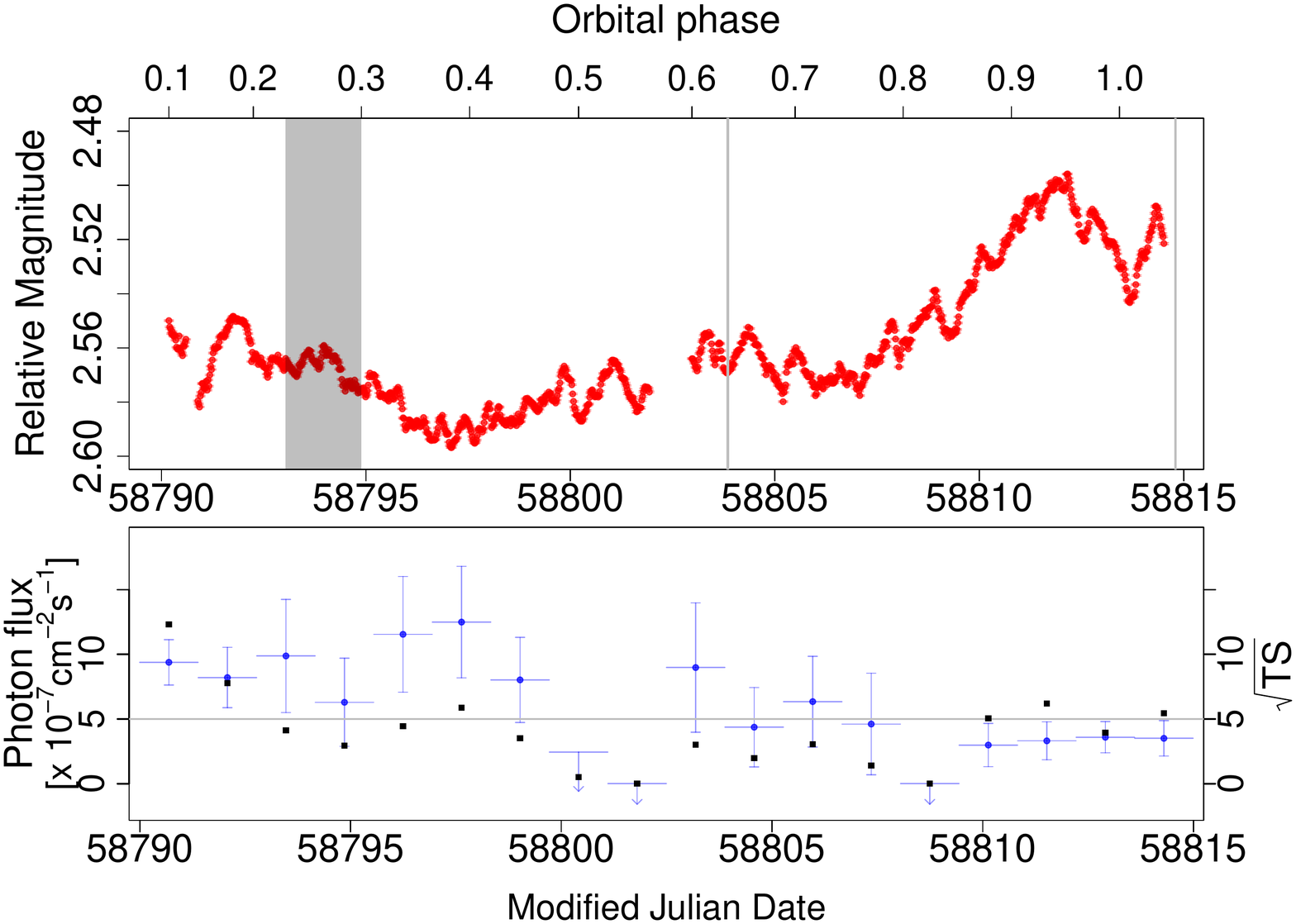}}
        %The \textit{TESS} optical light curve discussed in Section \ref{sec:optical} (relative to the average magnitude). The periastron has been located at phase [0.23-0.3] (noted in black shaded area) \textbf{que queremos decir con esta frase?}.
      \caption{Simultaneous optical and gamma ray light curves. 
      %\LEt{ Please provide a title for this figure.}
      Top panel: \textit{TESS} light curve of LS I +61 303 in November 2019
      with a binning time of half an hour. The vertical lines represent the dates when optical spectroscopy was obtained. 
      The black shaded area comprises the periastron phase, which has been located within the interval [0.23-0.3].
      The top horizontal axis gives the orbital phase of the data computed using the 26.496 d orbital period and phase origin on Julian date 2443366.775. 
      The vertical axis is plotted in terms of relative magnitude, computed as $-2.5 \log{(TESS~{\rm count~rate})}$ plus an arbitrary zero point,
      for easier comparison with historical photometric studies that use magnitudes instead of flux.
      Bottom panel: In blue points, the \emph{Fermi}-LAT photon flux of 4FGL J0240.5+6113 computed in bins of 32.9 hours in the same orbital period as the optical data (with only statistical errors). The black dots stand for the source detection significance ($\sqrt{\rm{TS}}$) in each time bin. The black horizontal line marks the $5\sigma$ detection significance.}
         \label{Fif:Fermilightcurve}
   \end{figure*}

The orbital modulation of LS I +61 303 is also seen in other wavelength domains. The optical variability, first reported by \citet{1986A&A...154L..30P},
was later shown by \citet{1989MNRAS.239..733M} to follow the orbital period. A similar behavior is present in
X-rays as well \citep{1997A&A...320L..25P}. The X-ray counterpart, originally detected with the {\it Einstein} satellite \citep{1981ApJ...247L..85B}, 
has been extensively observed at these energies (with multiple instruments, including {\it ROSAT}, {\it ASCA}, {\it RXTE}, {\it XMM-Newton}, 
{\it INTEGRAL}, {\it Swift/XRT}, and \emph{Chandra}; \citealt{1995A&A...299..751G,Leahy1997TheAX,2000ApJ...528..454H,Chernyakova:2006wx,
2006A&A...459..901S,2007A&A...474..575E,2007ApJ...664L..39P, 2012ApJ...744..106T, 2012ApJ...744L..13L}). X-ray flux variations have been reported on timescales of days \citep{1995A&A...299..751G,1996A&A...305..817T}  and  shorter \citep{2011ApJ...733...89L}. 
Via high-energy (HE) gamma rays, \emph{Fermi}-LAT \citep{2009ApJ...701L.123A} detected the modulated GeV emission of the source, revealing periodic outbursts slightly after the periastron passage ($\phi \sim 0.3-0.45$  \citealt{2012ApJ...749...54H}).

Finally, at very high energies, the source was first detected with MAGIC \citep{2006Sci...312.1771A} and later observed in different campaigns \citep{2008ApJ...679.1427A,MAGIC:2008eqb,2012ApJ...746...80A}. The TeV peak has been detected at phases close to the apastron ($\phi \sim 0.6-0.7$) at a flux level of up to $16\%$ of the Crab Nebula above 400 GeV.

In this work we report optical microflares on a daily timescale 
during an orbital period of LS I +61 303 with the \textit{Transiting Exoplanet Survey Satellite} (\textit{TESS}) and the search for their counterparts via
HE gamma rays with the \emph{Fermi} Large Area Telescope (LAT). 
We describe the optical observations in Sect. \ref{sec:optical} and summarize the LAT data analysis in Sect.\ 3. Finally, Sect. \ref{sec:conclusions} is 
devoted to a discussion and conclusions.

   \section{Optical  observations and analysis}
\label{sec:optical}

\subsection{Space-based photometry}
   
LS I +61 303 was included in the field of view of 
%the {\it Transiting Exoplanet Survey Satellite} 
\textit{TESS}\footnote{{\tt https://tess.mit.edu}} during almost a full orbital cycle in Sector 18 observed 
in November 2019. This space observatory provides continuous multi-target optical photometry in the  600-1000 nm range centered on the Cousins $I_C$ band at a wavelength of 786.5 nm.
With  an exposure time of 1426 s, {\it TESS} enables an unprecedented
sampling of the LS I +61 303 behavior at optical wavelengths.
%These data points belong to \textit{TESS} Sector 18 with exposure times of 1426 s that enable an unprecedented sampling of the LS I +61 303
%behavior at optical wavelengths.
 
The LS I +61 303 light curve was generated using the standard data reduction tools provided by the \textit{TESS} team \citep{2018ascl.soft12013L},
starting from the full frame image files. Special care was taken to make sure that effects of nearby stars did not affect the final result,
which is presented in the top panel of Fig. \ref{Fif:Fermilightcurve}.
The pixel level de-correlation technique \citep{2015ApJ...805..132D}  was also taken into account, but only minor differences were seen.
Remarkably, the resulting light curve shows both a long-term variability trend  over the orbital cycle and short-term, rapid ($\sim 1$ day) 
small-amplitude fluctuations of $\sim 0.01-0.02$ mag. 

\begin{figure}
   \resizebox{\hsize}{!}
        {\includegraphics[width=\linewidth]{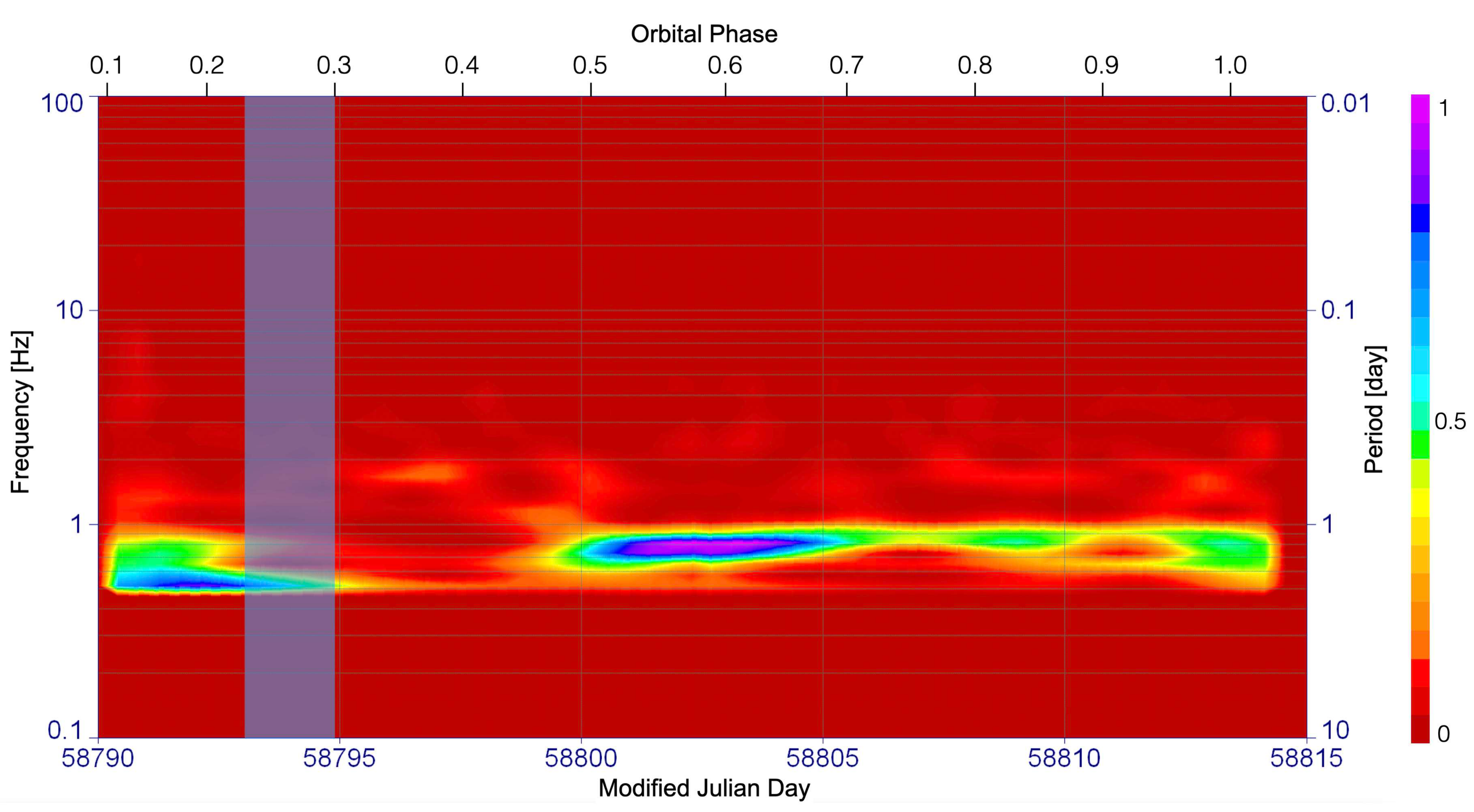}}
      \caption{Wavelet analysis of the LS I +61 303 light curve. The vertical shaded region represents the estimated periastron passage region.}
         \label{wavelets}
   \end{figure}

The optical microflares we report from the {\it TESS} data of LS I +61 303, detected on daily timescales, constitute a newly observed feature in the emission from 
this remarkable system. 
We used a continuous wavelet-based approach to detect temporal coherent patterns in our nonstationary observations \citep[see, for instance, ][]{burrus1998}. A Morlet mother wavelet with wavenumber 8 was chosen for a better compromise between time and frequency resolution. Only frequencies higher than 0.5 Hz were explored. The spectrogram that shows the mean squared amplitude power 
%(si lo prefieres en db me lo dices; db sería ~log (Re^2+Im^2), y lo otro sería ~(Re^2+Im^2), donde Re e Im son parte real e imaginaria de la wavelet en un tiempo y frecuencia dados). Evidentemente con db sale más información, pero guarrea un poco el gráfico) 
appears in Fig. \ref{wavelets} on a normalized scale. In it, 
a slowly evolving periodicity pattern with a recurrence interval of $\sim 1$ day can be seen.  
 Microflares appear to
repeat slightly faster  ($\sim 10$ \%) in the vicinity of periastron passage and slow down when approaching apastron.

\begin{figure}
   \resizebox{\hsize}{!}
        {\includegraphics[width=\linewidth, angle=-90]{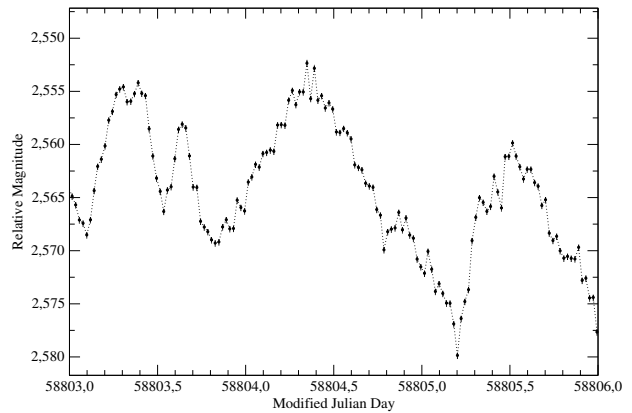}}
      \caption{Zoomed-in view of a sector of the {\it TESS} light curve, showing the shortest timescales of variability. Error bars are smaller than 1 millimagnitude.}
         \label{zoom}
   \end{figure}

 Concerning the shortest timescales of optical variability, in  Fig. \ref{zoom} it can be seen that some flare rises and decays occur on timescales of $\tau \sim 0.1$ days, which
 are interestingly reminiscent of previously observed fast radio variability \citep{1997A&A...328..283P}.
 Given the finite speed of light,  an upper limit to the size of the flaring region is  $c \tau  \sim 17$ AU.
 
\subsection{Ground-based spectroscopy}  \label{halfa}
  
\begin{figure}
\includegraphics[width=7.5cm, angle=-90]{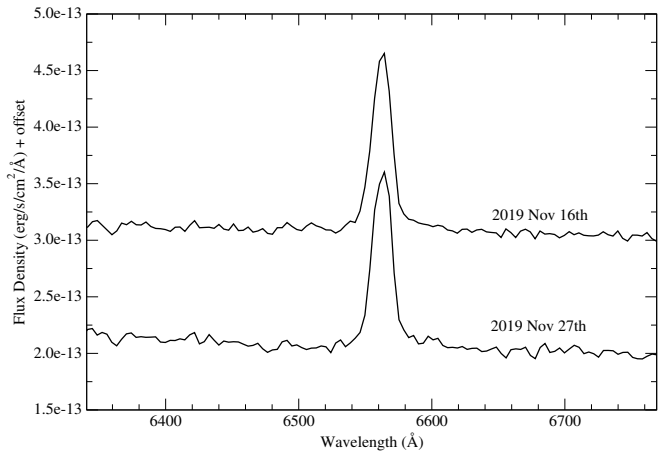}
%\imageii
\caption{   \label{sp_opt}  Optical spectra of LS I +61 303 contemporaneous to the TESS monitoring.
}%% no full stop at the end
\end{figure}

In parallel, low-resolution optical spectroscopy of LS I +61 303 was
fortunately
 obtained in coincidence with the \textit{TESS} monitoring using the 41 cm telescope at the
Observatory of the University of Ja\'en \citep{2017BlgAJ..26...91M}, equipped with a LISA 
spectrograph from Shelyak Instruments.
Data reduction and wavelength calibration using Ne lamps were performed by means of standard IRAF  
tools.
Spectra acquisition took place
on November 16 and 27, 2019 (MJD 58803.85 and 58814.80), and they are shown in Fig. \ref{sp_opt} around the H$\alpha$ feature.
The emission line flux remained nearly constant at the
$2.8 \times 10^{-12}$ erg s$^{-1}$ cm$^{-2}$ level with an average equivalent width of $-13.6$ \AA\ (estimated uncertainty 10\%).
 
%\begin{figure}
%\includegraphics[width=8.5cm, angle=0]{tess_mag.eps}
%\imageii
%\caption{   \label{tess_lc}  TESS light curve of LS I +61 303 during 2019 November. The vertical lines represent the dates
%when optical spectroscopy was obtained. The horizontal blue scale gives the orbital phase of the data computed using the 26.496 d orbital period
%and phase origin on Julian date 2443366.775. \textbf{A lo mejor esta figura y la 4 se pueden poner juntas...}
%}%% no full stop at the end
%\end{figure}

\section{Gamma-ray observations and analysis}
\label{sec:fermi}
%To search for the gamma-ray counterpart of the optical flares presented in the previous section, we aim at analyzing the \emph{Fermi}-LAT data towards LS I +61 303 recorded during the orbital period studied in optical wavelengths. In particular, we aim to obtain the gamma-ray light curve of LS I +61 303 along the cited period to explore its correlation with the optical light curve derived from TESS observations. We are first interested in determining the smallest timescales that guarantee a significant detection (of LS I +61 303) in every time bin of the \emph{Fermi}-LAT light curve. Hence, allowing us to compute an adequate time binning regarding the gamma-ray light curve of the source (according to its level of flux).

%We calculated the shortest binning in time appropriate to analyze the LAT data of LS I +61 303 by simulating \emph{Fermi} observational data (towards the direction of the source) with the tool \emph{gtobssim}. We followed the method described in Section 3.2 of \citet{2017ApJ...836...68T}. \emph{gtobssim} facilitates simulating point and diffuse source observations with \emph{Fermi} for a specific spectral shape and selected region of interest (ROI) both in the survey or pointed modes. We derived the source model employed to perform the simulations from five years of \emph{Fermi}-LAT data towards the position of LS I +61 303. 

We downloaded the {\it Fermi}-LAT \citep[P8R3,][]{Atwood:2013rka,bruel2018fermilat} data toward LS I +61 303 \citep[securely identified with the LAT gamma-ray source 4FGL J0240.5+6113;][]{2009ApJ...701L.123A} recorded from September 8, 2016, to September 8, 2021 (or 494995769 - 652762170 seconds in \emph{Fermi} mission elapsed time). This time interval fully overlaps with the {\it TESS} pointing.
The selected region of interest (ROI) is defined by a radius of $15\degr{}$ centered around the position of 4FGL J0240.5+6113 \citep[$\rm{RA}= 40.13\degr{}$ and $\rm{Dec}= 61.23\degr{}$; ][]{2009ApJ...701L.123A}. 
We analyzed only the dubbed SOURCE 
class events at energies from 100 MeV (to avoid events poorly reconstructed due to the large angular resolution) to 500 GeV, with a maximum zenith angle of 90\degr{} to eliminate Earth limb events.

\subsection{Getting an LS I +61 303 improved gamma-ray spectrum}

The model fitted to the LAT data described above includes all the LAT sources listed in the \emph{Fermi}-LAT Fourth Source Catalog \citep[4FGL;][]{Abdollahi_2020} within a $20\degr{}$  radius around the ROI center, plus the Galactic and extra-galactic diffuse gamma-ray components described with the latest available version of the Galactic (gll\_iem\_v07) and isotropic (iso\_P8R3\_SOURCE\_V3\_v1) diffuse emission models. The model's free parameters correspond to those of the sources within $5\degr{}$ of the position of 4FGL J0240.5+6113, the parameters regarding the isotropic and Galactic diffuse emission models, and the normalization parameter of all the sources in the model with a test statistic greater than ten ($TS > 10$). The test statistic is defined from the maximum likelihood function value over the ROI when including the source in the model ($L$) and if accounting only for background ($L_{0}$), that is, $TS = 2 \log{(L/L_{0})}$ \citep[see][]{1996ApJ...461..396M}. 
Therefore, the value of $\sqrt{TS}$ is usually interpreted as the detection significance of the source.

We analyzed the available LAT data  by means of a joint likelihood fitting process using the \textsc{fermipy} \textsc{python} package (version 1.0.1), which is based on the \textsc{Fermi Science Tools} \citep{2017ICRC...35..824W}. 
The response of the LAT instrument was evaluated with the instrument response functions (version P8R3\_SOURCE\_V2), applying the energy dispersion correction to all the sources in the model (except for the isotropic diffuse emission). We performed the analysis with the data binned in eight energy bins per decade and spatial bins of 0.1\degr{} in size.

We fitted the spectral energy distribution of the source with the \textsc{fermipy} extension method (which is based on a likelihood ratio test with respect to the point-source hypothesis 
to both a log-parabola ($dN/dE = N_{0} \times (E/E_{\rm ref})^{-(\alpha +\beta \log (E/E_{\rm ref}))}$) and a broken-power-law model ($dN/dE = N_{0} \times (E/E_{\rm break})^{-\Gamma}$, where $\Gamma = \Gamma_{1}$ for $E < E_{\rm break}$ and $\Gamma = \Gamma_{2}$ otherwise, using 13 energy bins (spanning from 100 MeV to 500 GeV) and the model of the whole ROI. Finally, we tested the consistency of the results by studying the systematic uncertainties.   
The uncertainties mainly stem from the LAT effective area ($\rm{A}_{eff}$) and the Galactic diffuse emission model. 
The systematic uncertainties regarding the LAT effective area are examined with the bracketing $\rm{A}_{eff}$ method\footnote{\url{https://fermi.gsfc.nasa.gov/ssc/data/analysis/scitools/Aeff_Systematics.html}}, and those regarding the diffuse Galactic model were estimated by artificially changing its normalization by $\pm 6$ per cent with respect to the best-fit one \citep[see, e.g., ][]{Ajello_2011,2018Jian}.

\begin{figure}
   \resizebox{\hsize}{!}
        {\includegraphics[width=\linewidth]{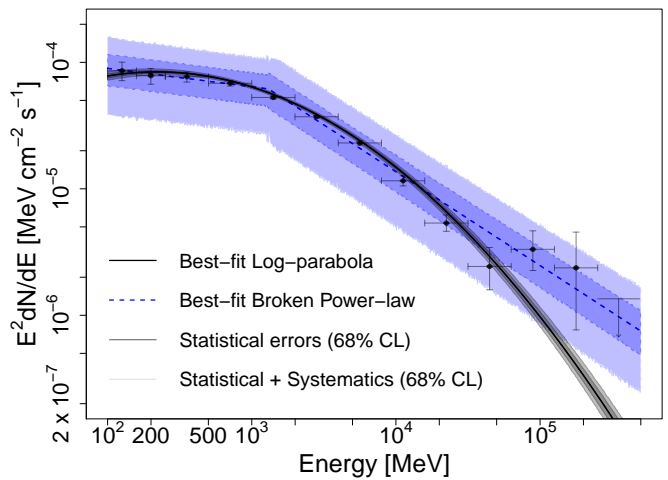}}
      \caption{Spectral energy distribution derived from \emph{Fermi}-LAT for LS I +61 303 (in black dots, accounting for systematical errors) with the best-fit  log-parabola (solid black line) and broken-power-law (dashed blue line) models. The dark and light shadowed regions correspond to the $1\sigma$ error of the best-fit models, i.e., with only statistical errors and accounting for systematical ones, respectively.}
         \label{Fif:FermiSED}
   \end{figure}

The LAT source (i.e., 4FGL J0240.5+6113) is detected at very high significance (with $\sqrt{TS} \approx 265$) as point-like at the position $\rm{RA} = 40.143\degr{} \pm 0.002\degr{}$ and $\rm{Dec} = 61.235\degr{} \pm 0.002\degr{}$ (with only statistical errors). 
The spectral parameters of the best-fit log-parabola model correspond to $N_{0} = (4.31 \pm 0.04_{\rm stat} \pm 0.12_{\rm sys}) \times 10^{-11}$ MeV$^{-1}$ cm$^{-2}$ s$^{-1}$, $\alpha = 2.40 \pm 0.01_{\rm stat} \pm 0.02_{\rm sys}$, and $\beta = 0.119 \pm 0.004_{\rm stat} \pm 0.006_{\rm sys}$ (referenced to an energy of $1.18$ GeV).
The best-fit parameters for a broken power law consist of $N_{0} = (3.61 \pm 0.73_{\rm stat} \pm 0.99_{\rm sys}) \times 10^{-11}$ MeV$^{-1}$ cm$^{-2}$ s$^{-1}$, $\Gamma_{1} = 2.146 \pm 0.009_{\rm stat} \pm 0.075_{\rm sys}$, $\Gamma_{2} = 2.74 \pm 0.04_{\rm stat} \pm 0.07_{\rm sys}$, and $E_{\rm break} = 1307 \pm 117_{\rm stat} \pm 304_{\rm sys}$ MeV  (see both fitted models in Fig. \ref{Fif:FermiSED}).
%{\bf The Test Statistic favors the log-parabola hypothesis over the broken power-law one with $\sqrt{\Delta TS} \approx 7.3$. 
%However, the broken power-law shape best characterizes the emission at the highest energies ($E \gtrsim 100$ GeV).}
The test statistic seems to favor the log-parabola hypothesis over the broken-power-law one with $\sqrt{\Delta TS} \approx 7.3$.
We note, however, that the two hypotheses are not nested and therefore the result of the likelihood ratio test cannot be trusted quantitatively.
We searched for a possible extension of the source (with the \emph{fermipy extension} method), but considering the source as extended does not result in better modeling of the data (i.e., it does not significantly improve the likelihood with respect to the point-like case,  with $TS_{\rm extended} - TS_{\rm point-like} < 25$  for both a 2D Gaussian source and a radial disk morphological model. 

\subsection{Assessing the shortest gamma-ray time bin feasible for LS I +61 303}

The best-fit broken-power-law model summarized above was then used to perform  \emph{Fermi}-LAT observation simulations regarding 4FGL J0240.5+6113 with different observation times, ranging from 1 to 72 hours (in 16 steps).
We performed the simulations with the \emph{Fermi} tool {\sc gtobssim}, which allowed us to simulate (non-variable) point-like sources characterized by a specific spectral shape, either a power-law or broken-power-law spectrum. {\sc gtobssim} does not implement the log-parabola form, but the source's spectrum exhibits a noticeable curvature (see Fig. \ref{Fif:FermiSED}). Hence, we chose the broken-power-law model.
The simulated ROI was the same as the one for the LAT data analysis of 4FGL J0240.5+6113 (as described above). The observation simulations only account for the emission from the source of interest and for the Galactic and isotropic diffuse emission models. We also assumed that the flux from 4FGL J0240.5+6113 is steady during the simulated observation time. The simulations were next analyzed as real LAT data, following the same steps as performed before to find the best-fit position and spectral parameters of the LAT source. A simple linear fit of the source detection significance (approximated as $\sqrt{TS}$) with respect to the observation time yields an average time for achieving the $5\sigma$ detection of $32.9 \pm 4.7$ hours (the linear fit being a reduced chi-square of $\chi^{2}_{dof = 14} = 1.6$).

The results from the simulations imply that under the assumptions explained above, the source would be detectable at a $5\sigma$ significance for intervals of $\sim 33$ hours. However, we would need larger time bins to properly derive the {\it Fermi} light curve of LS I +61 303. Otherwise, the detection of the source would not reach the $5\sigma$ significance in all (or most) of the time bins, firstly because of the presence of other gamma-ray sources in the ROI and secondly because the flux from LS I +61 303 is clearly not steady on timescales of days \citep[we note the orbital modulation of the flux with a $\sim 26.5$ day periodicity;][]{2009ApJ...701L.123A}. Hence, this result rules out the possibility of detecting the intra-orbit variability of LS I +61 303 in the LAT data on daily timescales regarding the orbital period studied in optical wavelengths, preventing us from searching the gamma-ray counterparts of the detected optical microflares. The lower panel of
Fig. \ref{Fif:Fermilightcurve} shows the light curve of 4FGL J0240.5+6113 during the same orbital period as studied in the optical and with the binning derived from the performed  simulations. We note that the average detection significance of the source among the different time bins was close to $5\sigma$ (i.e., $\approx (4.5 \pm 2.7)\sigma$, as expected from the simulations).

%\section{Concurrent optical spectroscopy}

\section{Discussion and conclusions} \label{discu}
\label{sec:conclusions}

A multiwavelength approach to the optical microflare phenomenon, during the {\it TESS} Sector 18 observations,
 can only {currently} be  attempted in the gamma-ray domain using {\it Fermi}-LAT data as no other 
continuous and simultaneous monitoring is available. Hopefully, future coordinated observations will enable this approach in the radio or X-rays.
Concurrent optical and gamma-ray microflares in LS I +61 303 would imply a common origin, most likely in shocks \citep{2018MNRAS.479..687S,2020NatAs...4..776A}. 
This expectation is inspired by 
semi-analytic models of nova shocks; these models show that most of the shocks' kinetic energy that is dissipated while they move through a dense medium is radiated as thermal X-rays, which can be absorbed by neutral gas ahead or behind the shock and reemitted in optical frequencies 
\citep{2014MNRAS.442..713M}. \cite{2015MNRAS.450.2739M} argue that a shock in which gamma-ray emission is radiated and observable (i.e., not absorbed) necessarily radiates the bulk of its dissipated thermal energy in the optical, while gamma-ray absorption at early times may introduce a delay in the gamma-ray emission. Hence, the optical emission we observed could be shock-powered, even if the gamma-ray emission is suppressed.  In this context, the shock regions are naturally expected to form while a compact object magnetosphere interacts with the Be star circumstellar disk.
Their size is expected to exceed the orbital dimensions and reach a few AU in the LS I +61 303 case 
(in agreement with the previous $\sim 17$ AU upper limit.).
Moreover, the 
decretion disk was  well developed at the time of {\it TESS} observations, as the measured H$\alpha$ equivalent width (given in Sect. \ref{halfa})
was closer to the maximum reported values \citep{2014A&A...561L...2Z}. 
However, how the disk structure could relate to the apparent $\sim 1$ d 
recurrence interval  
is not a straightforward issue and likely depends on the density pattern being sampled by the compact companion.
  
Based on the above nova-analogy scenario, we searched for the possible HE gamma-ray counterparts of the microflares in LAT data.
%Unfortunately,  intra-orbit variability on daily timescales in LAT data cannot be detected through simulations given the too low flux level of the source.  
Unfortunately, the possibility of detecting  intra-orbit variability on daily timescales in LAT data is ruled out through simulations given the too low flux level of the source.
Nevertheless, its expected value can be crudely estimated as follows.
According to \cite{2015MNRAS.450.2739M}, in the shock-powered scenario for optical microflares, we could expect gamma-ray counterparts with luminosity $L_{\gamma} \lesssim \epsilon_{\gamma} \epsilon_{\rm nth} L_{\rm opt}$. Here,  $L_{\rm opt}$ is the optical luminosity of the flares, $\epsilon_{\rm nth}$ is the fraction of the total power dissipated by shocks that goes into accelerating nonthermal ions or electrons, and $\epsilon_{\gamma}$ includes the radiative efficiency in the accelerated particles and the total gamma-ray emission radiated in the LAT bandpass. 
On the one hand, the optical luminosity of the observed microflares results in $L_{\rm opt} \sim 2.1 \times 10^{34}$ erg/s for fluctuations of $\sim 0.02$ magnitude over an average magnitude for LS I +61 303 of 9.55 in the I band (\citealt{2015A&A...574A..33R}, approximately centered in the same wavelength as the {\it TESS} observations), and a distance of 2.63 kpc (Gaia EDR3, \citealt{2021A&A...649A...1G}). On the other hand, the next upper limit, $\epsilon_{\gamma} \epsilon_{\rm nth} \lesssim 0.03$, holds according to nonrelativistic shock simulations (both for hadronic and leptonic models, again following \citealt{2015MNRAS.450.2739M}). 
This implies a gamma-ray luminosity for the flares of $L_{\gamma} \sim 6.3 \times 10^{32}$ erg/s, or a photon flux at 2.63 kpc on the order of $7.6 \times 10^{-13}$ erg cm$^{-2}$ s$^{-1}$ (taking an average photon energy in the LAT passband of $E_{\rm break} = 1307$ MeV). The predicted photon flux is thus far below (by about three orders of magnitudes) the sensitivity achieved in the LAT gamma-ray light curve of LS I +61 303 (in bins of 33 hours; see the lower panel of Fig. \ref{Fif:Fermilightcurve}).
Although we believe that  the shock-driven scenario is feasible, it remains hard to prove observationally.

Other alternative scenarios are worth considering.
In particular, one could speculate that the observed low-amplitude variability is of rotation or pulsation origin.
Classical Be stars are physically understood as rapidly rotating B stars. However, in this case, the reported optical variability  can hardly be related to a 
strictly periodic rotation or pulsation because no clearly constant period is apparent in the data wavelet analysis or other periodograms. Finally, if LS I +61 303 hosts a black hole surrounded by an
accretion disk, as proposed by \cite{2017MNRAS.468.3689M}, the reported light curve could be understood in terms of flickering. Nevertheless, this type of variability in active binaries is usually observed on shorter timescales (of around an hour), and the black hole assumption appears to have recently been ruled out.

To conclude, mini-flares in LS I +61 303 are newly observed phenomena, our understanding of which remains challenging as we await future data and theoretical work.

%In addition, rotation should be broadening the lines in the spectra \textbf{(can we see it? give an upper limit?)}. Non-periodic pulsations originated in the corona of the star have timescales much shorter. \textbf{Thus, bolometric luminosity could be originating as X-rays from shocks between the neutron star magnetosphere and the Be decretion disc, that are then absorbed and reprocessed to emerge in the optical. (Â¿y si el sistema fuese un agujero negro? Â¿se podria explicar con choques?)} 

\begin{acknowledgements}
EM acknowledges support by grant P18-FR-1580 from the  
Consejer\'{\i}a de Econom\'{\i}a y Conocimiento de la Junta de Andaluc\'{\i}a 
under the Programa Operativo FEDER 2014-2020.
ESA, PLLE and JM acknowledge support  by grant PID2019-105510GB-C32 / AEI / 10.13039/501100011033 from 
State Agency for Research of the Spanish Ministry of Science and Innovation 
 entitled {\it High energy sources with outflows at different scales: observation of galactic sources}. We also acknowledge support
 by Consejer\'{\i}a de Econom\'{\i}a, Innovaci\'on, Ciencia y Empleo of Junta de Andaluc\'{\i}a as research group FQM- 322, as well as FEDER funds.
 JMP, DdS and JN acknowledge financial support from the State Agency for Research of the Spanish Ministry of Science and Innovation
under grant PID2019-105510GB-C31 and through the Unit of Excellence Mar\'{\i}a de Maeztu 2020-2023 award to the
Institute of Cosmos Sciences (CEX2019-000918-M).%, and by the Catalan DEC grant 2017 SGR 643.
 This paper includes data collected with the TESS mission, obtained from the MAST data archive at the Space Telescope Science Institute (STScI). Funding for the TESS mission is provided by the NASA Explorer Program. STScI is operated by the Association of Universities for Research in Astronomy, Inc., under NASA contract NAS 5-26555.\end{acknowledgements}

\bibliographystyle{aa} % style aa.bst
\bibliography{61303} % your references Yourfile.bib

\begin{thebibliography}{65}
\expandafter\ifx\csname natexlab\endcsname\relax\def\natexlab#1{#1}\fi

\bibitem[{{Abdo} {et~al.}(2009){Abdo}, {Ackermann}, {Ajello}, {Atwood},
  {Axelsson}, {Baldini}, {Ballet}, {Barbiellini}, {Bastieri}, {Baughman},
  {Bechtol}, {Bellazzini}, {Berenji}, {Blandford}, {Bloom}, {Bonamente},
  {Borgland}, {Bregeon}, {Brez}, {Brigida}, {Bruel}, {Burnett}, {Caliandro},
  {Cameron}, {Caraveo}, {Casandjian}, {Cavazzuti}, {Cecchi}, {{\c{C}}elik},
  {Charles}, {Chaty}, {Chekhtman}, {Cheung}, {Chiang}, {Ciprini}, {Claus},
  {Cohen-Tanugi}, {Cominsky}, {Conrad}, {Corbel}, {Corbet}, {Cutini}, {Dermer},
  {de Angelis}, {de Luca}, {de Palma}, {Digel}, {Dormody}, {do Couto e Silva},
  {Drell}, {Dubois}, {Dubus}, {Dumora}, {Farnier}, {Favuzzi}, {Fegan}, {Focke},
  {Frailis}, {Fukazawa}, {Funk}, {Fusco}, {Gargano}, {Gasparrini}, {Gehrels},
  {Germani}, {Giebels}, {Giglietto}, {Giordano}, {Glanzman}, {Godfrey},
  {Grenier}, {Grondin}, {Grove}, {Guillemot}, {Guiriec}, {Hanabata}, {Harding},
  {Hayashida}, {Hays}, {Hill}, {Hughes}, {J{\'o}hannesson}, {Johnson},
  {Johnson}, {Johnson}, {Johnson}, {Kamae}, {Katagiri}, {Kataoka}, {Kawai},
  {Kerr}, {Kn{\"o}dlseder}, {Kocian}, {Kuehn}, {Kuss}, {Lande}, {Larsson},
  {Latronico}, {Longo}, {Loparco}, {Lott}, {Lovellette}, {Lubrano}, {Madejski},
  {Makeev}, {Marelli}, {Mazziotta}, {McEnery}, {Meurer}, {Michelson},
  {Mitthumsiri}, {Mizuno}, {Monte}, {Monzani}, {Morselli}, {Moskalenko},
  {Murgia}, {Nolan}, {Nuss}, {Ohsugi}, {Okumura}, {Omodei}, {Orlando}, {Ormes},
  {Paneque}, {Panetta}, {Parent}, {Pelassa}, {Pepe}, {Pesce-Rollins}, {Piron},
  {Porter}, {Rain{\`o}}, {Rando}, {Ray}, {Razzano}, {Rea}, {Reimer}, {Reimer},
  {Reposeur}, {Ritz}, {Rochester}, {Rodriguez}, {Romani}, {Ryde},
  {Sadrozinski}, {Sanchez}, {Sander}, {Saz Parkinson}, {Scargle}, {Sgr{\`o}},
  {Shaw}, {Sierpowska-Bartosik}, {Siskind}, {Smith}, {Smith}, {Spandre},
  {Spinelli}, {Striani}, {Strickman}, {Suson}, {Tajima}, {Takahashi},
  {Takahashi}, {Tanaka}, {Thayer}, {Thayer}, {Thompson}, {Tibaldo}, {Torres},
  {Tosti}, {Tramacere}, {Uchiyama}, {Usher}, {Vasileiou}, {Vilchez}, {Vitale},
  {Waite}, {Wang}, {Winer}, {Wood}, {Ylinen}, \&
  {Ziegler}}]{2009ApJ...701L.123A}
{Abdo}, A.~A., {Ackermann}, M., {Ajello}, M., {et~al.} 2009, \apjl, 701, L123

\bibitem[{Abdollahi {et~al.}(2020)Abdollahi, Acero, Ackermann, Ajello, Atwood,
  Axelsson, Baldini, Ballet, Barbiellini, Bastieri, Gonzalez, Bellazzini,
  Berretta, Bissaldi, Blandford, Bloom, Bonino, Bottacini, Brandt, Bregeon,
  Bruel, Buehler, Burnett, Buson, Cameron, Caputo, Caraveo, Casandjian, Castro,
  Cavazzuti, Charles, Chaty, Chen, Cheung, Chiaro, Ciprini, Cohen-Tanugi,
  Cominsky, Coronado-Bl{\'{a}}zquez, Costantin, Cuoco, Cutini, D'Ammando,
  DeKlotz, de~la Torre~Luque, de~Palma, Desai, Digel, Lalla, Mauro, Venere,
  Dom{\'{\i}}nguez, Dumora, Dirirsa, Fegan, Ferrara, Franckowiak, Fukazawa,
  Funk, Fusco, Gargano, Gasparrini, Giglietto, Giommi, Giordano, Giroletti,
  Glanzman, Green, Grenier, Griffin, Grondin, Grove, Guiriec, Harding, Hayashi,
  Hays, Hewitt, Horan, J{\'{o}}hannesson, Johnson, Kamae, Kerr, Kocevski,
  Kovac'evic', Kuss, Landriu, Larsson, Latronico, Lemoine-Goumard, Li,
  Liodakis, Longo, Loparco, Lott, Lovellette, Lubrano, Madejski, Maldera,
  Malyshev, Manfreda, Marchesini, Marcotulli, Mart{\'{\i}}-Devesa, Martin,
  Massaro, Mazziotta, McEnery, Mereu, Meyer, Michelson, Mirabal, Mizuno,
  Monzani, Morselli, Moskalenko, Negro, Nuss, Ojha, Omodei, Orienti, Orlando,
  Ormes, Palatiello, Paliya, Paneque, Pei, Pe{\~{n}}a-Herazo, Perkins, Persic,
  Pesce-Rollins, Petrosian, Petrov, Piron, Poon, Porter, Principe, Rain{\`{o}},
  Rando, Razzano, Razzaque, Reimer, Reimer, Remy, Reposeur, Romani, Parkinson,
  Schinzel, Serini, Sgr{\`{o}}, Siskind, Smith, Spandre, Spinelli, Strong,
  Suson, Tajima, Takahashi, Tak, Thayer, Thompson, Tibaldo, Torres, Torresi,
  Valverde, Klaveren, van Zyl, Wood, Yassine, \& Zaharijas}]{Abdollahi_2020}
Abdollahi, S., Acero, F., Ackermann, M., {et~al.} 2020, The Astrophysical
  Journal Supplement Series, 247, 33

\bibitem[{{Acciari} {et~al.}(2008){Acciari}, {Beilicke}, {Blaylock},
  {Bradbury}, {Buckley}, {Bugaev}, {Butt}, {Byrum}, {Celik}, {Cesarini},
  {Ciupik}, {Chow}, {Cogan}, {Colin}, {Cui}, {Daniel}, {Duke}, {Ergin},
  {Falcone}, {Fegan}, {Finley}, {Fortin}, {Fortson}, {Gall}, {Gibbs},
  {Gillanders}, {Grube}, {Guenette}, {Hanna}, {Hays}, {Holder}, {Horan},
  {Hughes}, {Hui}, {Humensky}, {Kaaret}, {Kieda}, {Kildea}, {Konopelko},
  {Krawczynski}, {Krennrich}, {Lang}, {LeBohec}, {Lee}, {Maier}, {McCann},
  {McCutcheon}, {Millis}, {Moriarty}, {Mukherjee}, {Nagai}, {Ong}, {Pandel},
  {Perkins}, {Pizlo}, {Pohl}, {Quinn}, {Ragan}, {Reynolds}, {Rose},
  {Schroedter}, {Sembroski}, {Smith}, {Steele}, {Swordy}, {Toner}, {Valcarcel},
  {Vassiliev}, {Wagner}, {Wakely}, {Ward}, {Weekes}, {Weinstein}, {White},
  {Williams}, {Wissel}, {Wood}, \& {Zitzer}}]{2008ApJ...679.1427A}
{Acciari}, V.~A., {Beilicke}, M., {Blaylock}, G., {et~al.} 2008, \apj, 679,
  1427

\bibitem[{Ajello {et~al.}(2011)Ajello, Allafort, Baldini, Ballet, Barbiellini,
  Bastieri, Bechtol, Bellazzini, Berenji, Blandford, Bloom, Bonamente,
  Borgland, Bregeon, Brigida, Bruel, Buehler, Buson, Caliandro, Cameron,
  Caraveo, Casandjian, Cecchi, Charles, Chekhtman, Ciprini, Claus,
  Cohen-Tanugi, Cutini, de~Angelis, de~Palma, Dermer, do~Couto~e Silva, Drell,
  Drlica-Wagner, Dubois, Favuzzi, Fegan, Ferrara, Focke, Frailis, Fukazawa,
  Fukui, Fusco, Gargano, Gasparrini, Germani, Giglietto, Giommi, Giordano,
  Giroletti, Glanzman, Godfrey, Grove, Guiriec, Hadasch, Hanabata, Harding,
  Hayashi, Hays, Itoh, J{\'{o}}hannesson, Johnson, Kamae, Katagiri, Kataoka,
  Knödlseder, Kubo, Kuss, Lande, Latronico, Lee, Lionetto, Longo, Loparco,
  Lovellette, Lubrano, Mazziotta, Mehault, Michelson, Mizuno, Moiseev, Monte,
  Monzani, Morselli, Moskalenko, Murgia, Nakamori, Naumann-Godo, Nishino,
  Nolan, Norris, Nuss, Ohno, Ohsugi, Okumura, Omodei, Orlando, Ormes, Paneque,
  Parent, Pelassa, Pesce-Rollins, Pierbattista, Piron, Porter, Rain{\`{o}},
  Rando, Reimer, Reimer, Reposeur, Roth, Sadrozinski, Sgr{\`{o}}, Siskind,
  Smith, Spandre, Spinelli, Suson, Tajima, Takahashi, Tanaka, Thayer, Thayer,
  Tibaldo, Tibolla, Torres, Tosti, Tramacere, Troja, Uchiyama, Uehara, Usher,
  Vandenbroucke, Etten, Vasileiou, Vianello, Vilchez, Vitale, Waite, Wang,
  Winer, Wood, Yamamoto, Yamazaki, Yang, Yasuda, Ziegler, \&
  Zimmer}]{Ajello_2011}
Ajello, M., Allafort, A., Baldini, L., {et~al.} 2011, The Astrophysical
  Journal, 744, 80

\bibitem[{{Albert} {et~al.}(2006){Albert}, {Aliu}, {Anderhub}, {Antoranz},
  {Armada}, {Asensio}, {Baixeras}, {Barrio}, {Bartelt}, {Bartko}, {Bastieri},
  {Bavikadi}, {Bednarek}, {Berger}, {Bigongiari}, {Biland}, {Bisesi}, {Bock},
  {Bordas}, {Bosch-Ramon}, {Bretz}, {Britvitch}, {Camara}, {Carmona},
  {Chilingarian}, {Ciprini}, {Coarasa}, {Commichau}, {Contreras}, {Cortina},
  {Curtef}, {Danielyan}, {Dazzi}, {De Angelis}, {de los Reyes}, {De Lotto},
  {Domingo-Santamar{\'\i}a}, {Dorner}, {Doro}, {Errando}, {Fagiolini},
  {Ferenc}, {Fern{\'a}ndez}, {Firpo}, {Flix}, {Fonseca}, {Font}, {Fuchs},
  {Galante}, {Garczarczyk}, {Gaug}, {Giller}, {Goebel}, {Hakobyan},
  {Hayashida}, {Hengstebeck}, {H{\"o}hne}, {Hose}, {Hsu}, {Isar}, {Jacon},
  {Kalekin}, {Kosyra}, {Kranich}, {Laatiaoui}, {Laille}, {Lenisa}, {Liebing},
  {Lindfors}, {Lombardi}, {Longo}, {L{\'o}pez}, {L{\'o}pez}, {Lorenz},
  {Lucarelli}, {Majumdar}, {Maneva}, {Mannheim}, {Mansutti}, {Mariotti},
  {Mart{\'\i}nez}, {Mase}, {Mazin}, {Merck}, {Meucci}, {Meyer}, {Miranda},
  {Mirzoyan}, {Mizobuchi}, {Moralejo}, {Nilsson}, {O{\~n}a-Wilhelmi},
  {Ordu{\~n}a}, {Otte}, {Oya}, {Paneque}, {Paoletti}, {Paredes}, {Pasanen},
  {Pascoli}, {Pauss}, {Pavel}, {Pegna}, {Persic}, {Peruzzo}, {Piccioli},
  {Poller}, {Pooley}, {Prandini}, {Raymers}, {Rhode}, {Rib{\'o}}, {Rico},
  {Riegel}, {Rissi}, {Robert}, {Romero}, {R{\"u}gamer}, {Saggion},
  {S{\'a}nchez}, {Sartori}, {Scalzotto}, {Scapin}, {Schmitt}, {Schweizer},
  {Shayduk}, {Shinozaki}, {Shore}, {Sidro}, {Sillanp{\"a}{\"a}}, {Sobczynska},
  {Stamerra}, {Stark}, {Takalo}, {Temnikov}, {Tescaro}, {Teshima}, {Tonello},
  {Torres}, {Torres}, {Turini}, {Vankov}, {Vitale}, {Wagner}, {Wibig},
  {Wittek}, {Zanin}, \& {Zapatero}}]{2006Sci...312.1771A}
{Albert}, J., {Aliu}, E., {Anderhub}, H., {et~al.} 2006, Science, 312, 1771

\bibitem[{Albert {et~al.}(2009)}]{MAGIC:2008eqb}
Albert, J. {et~al.} 2009, Astrophys. J., 693, 303

\bibitem[{{Aleksi{\'c}} {et~al.}(2012){Aleksi{\'c}}, {Alvarez}, {Antonelli},
  {Antoranz}, {Asensio}, {Backes}, {Barrio}, {Bastieri}, {Becerra
  Gonz{\'a}lez}, {Bednarek}, {Berdyugin}, {Berger}, {Bernardini}, {Biland},
  {Blanch}, {Bock}, {Boller}, {Bonnoli}, {Borla Tridon}, {Bosch-Ramon},
  {Braun}, {Bretz}, {Ca{\~n}ellas}, {Carmona}, {Carosi}, {Colin}, {Colombo},
  {Contreras}, {Cortina}, {Cossio}, {Covino}, {Dazzi}, {De Angelis}, {De
  Caneva}, {De Cea del Pozo}, {De Lotto}, {Delgado Mendez}, {Diago Ortega},
  {Doert}, {Dom{\'\i}nguez}, {Dominis Prester}, {Dorner}, {Doro}, {Elsaesser},
  {Ferenc}, {Fonseca}, {Font}, {Fruck}, {Garc{\'\i}a L{\'o}pez}, {Garczarczyk},
  {Garrido}, {Giavitto}, {Godinovi{\'c}}, {Hadasch}, {H{\"a}fner}, {Herrero},
  {Hildebrand}, {H{\"o}hne-M{\"o}nch}, {Hose}, {Hrupec}, {Huber}, {Jogler},
  {Kellermann}, {Klepser}, {Kr{\"a}henb{\"u}hl}, {Krause}, {La Barbera},
  {Lelas}, {Leonardo}, {Lindfors}, {Lombardi}, {L{\'o}pez}, {L{\'o}pez},
  {Lorenz}, {Makariev}, {Maneva}, {Mankuzhiyil}, {Mannheim}, {Maraschi},
  {Mariotti}, {Mart{\'\i}nez}, {Mazin}, {Meucci}, {Miranda}, {Mirzoyan},
  {Miyamoto}, {Mold{\'o}n}, {Moralejo}, {Munar-Adrover}, {Nieto}, {Nilsson},
  {Orito}, {Oya}, {Paneque}, {Paoletti}, {Pardo}, {Paredes}, {Partini},
  {Pasanen}, {Pauss}, {Perez-Torres}, {Persic}, {Peruzzo}, {Pilia}, {Pochon},
  {Prada}, {Prada Moroni}, {Prandini}, {Puljak}, {Reichardt}, {Reinthal},
  {Rhode}, {Rib{\'o}}, {Rico}, {R{\"u}gamer}, {Saggion}, {Saito}, {Saito},
  {Salvati}, {Satalecka}, {Scalzotto}, {Scapin}, {Schultz}, {Schweizer},
  {Shayduk}, {Shore}, {Sillanp{\"a}{\"a}}, {Sitarek}, {Sobczynska}, {Spanier},
  {Spiro}, {Stamerra}, {Steinke}, {Storz}, {Strah}, {Suri{\'c}}, {Takalo},
  {Takami}, {Tavecchio}, {Temnikov}, {Terzi{\'c}}, {Tescaro}, {Teshima},
  {Tibolla}, {Torres}, {Treves}, {Uellenbeck}, {Vankov}, {Vogler}, {Wagner},
  {Weitzel}, {Zabalza}, {Zandanel}, \& {Zanin}}]{2012ApJ...746...80A}
{Aleksi{\'c}}, J., {Alvarez}, E.~A., {Antonelli}, L.~A., {et~al.} 2012, \apj,
  746, 80

\bibitem[{Atwood {et~al.}(2013)}]{Atwood:2013rka}
Atwood, W. {et~al.} 2013, in {Pass 8: Toward the Full Realization of the
  Fermi-LAT Scientific Potential}

\bibitem[{{Aydi} {et~al.}(2020){Aydi}, {Sokolovsky}, {Chomiuk}, {Steinberg},
  {Li}, {Vurm}, {Metzger}, {Strader}, {Mukai}, {Pejcha}, {Shen}, {Wade},
  {Kuschnig}, {Moffat}, {Pablo}, {Pigulski}, {Popowicz}, {Weiss}, {Zwintz},
  {Izzo}, {Pollard}, {Handler}, {Ryder}, {Filipovi{\'c}}, {Alsaberi},
  {Manojlovi{\'c}}, {Lopes de Oliveira}, {Walter}, {Vallely}, {Buckley},
  {Brown}, {Harvey}, {Kawash}, {Kniazev}, {Kochanek}, {Linford},
  {Mikolajewska}, {Molaro}, {Orio}, {Page}, {Shappee}, \&
  {Sokoloski}}]{2020NatAs...4..776A}
{Aydi}, E., {Sokolovsky}, K.~V., {Chomiuk}, L., {et~al.} 2020, Nature
  Astronomy, 4, 776

\bibitem[{Bednarek(2006{\natexlab{a}})}]{bednarek2006inverse}
Bednarek, W. 2006{\natexlab{a}}, Monthly Notices of the Royal Astronomical
  Society, 371, 1737

\bibitem[{Bednarek(2006{\natexlab{b}})}]{Bednarek:2006nc}
Bednarek, W. 2006{\natexlab{b}}, Mon. Not. Roy. Astron. Soc., 368, 579

\bibitem[{{Bignami} {et~al.}(1981){Bignami}, {Caraveo}, {Lamb}, {Markert}, \&
  {Paul}}]{1981ApJ...247L..85B}
{Bignami}, G.~F., {Caraveo}, P.~A., {Lamb}, R.~C., {Markert}, T.~H., \& {Paul},
  J.~A. 1981, \apjl, 247, L85

\bibitem[{{Bosch-Ramon} {et~al.}(2006){Bosch-Ramon}, {Paredes}, {Romero}, \&
  {Rib{\'o}}}]{2006A&A...459L..25B}
{Bosch-Ramon}, V., {Paredes}, J.~M., {Romero}, G.~E., \& {Rib{\'o}}, M. 2006,
  \aap, 459, L25

\bibitem[{Bruel {et~al.}(2018)Bruel, Burnett, Digel, Johannesson, Omodei, \&
  Wood}]{bruel2018fermilat}
Bruel, P., Burnett, T.~H., Digel, S.~W., {et~al.} 2018, Fermi-LAT improved
  Pass~8 event selection

\bibitem[{Burrus {et~al.}(1998)Burrus, Gopinath, \& Guo}]{burrus1998}
Burrus, C.~S., Gopinath, R., \& Guo, H. 1998, Introduction to Wavelets and
  Wavelet Transforms (New Jersey (USA): Prentice-Hall)

\bibitem[{{Casares} {et~al.}(2005){Casares}, {Ribas}, {Paredes}, {Mart{\'\i}},
  \& {Allende Prieto}}]{2005MNRAS.360.1105C}
{Casares}, J., {Ribas}, I., {Paredes}, J.~M., {Mart{\'\i}}, J., \& {Allende
  Prieto}, C. 2005, \mnras, 360, 1105

\bibitem[{Chernyakova {et~al.}(2006)Chernyakova, Neronov, \&
  Walter}]{Chernyakova:2006wx}
Chernyakova, M., Neronov, A., \& Walter, R. 2006, Mon. Not. Roy. Astron. Soc.,
  372, 1585

\bibitem[{{Deming} {et~al.}(2015){Deming}, {Knutson}, {Kammer}, {Fulton},
  {Ingalls}, {Carey}, {Burrows}, {Fortney}, {Todorov}, {Agol}, {Cowan},
  {Desert}, {Fraine}, {Langton}, {Morley}, \& {Showman}}]{2015ApJ...805..132D}
{Deming}, D., {Knutson}, H., {Kammer}, J., {et~al.} 2015, \apj, 805, 132

\bibitem[{Dermer \& B{\"o}ttcher(2006)}]{dermer2006gamma}
Dermer, C.~D. \& B{\"o}ttcher, M. 2006, The Astrophysical Journal, 643, 1081

\bibitem[{{Dubus}(2006)}]{2006A&A...456..801D}
{Dubus}, G. 2006, \aap, 456, 801

\bibitem[{{Esposito} {et~al.}(2007){Esposito}, {Caraveo}, {Pellizzoni}, {de
  Luca}, {Gehrels}, \& {Marelli}}]{2007A&A...474..575E}
{Esposito}, P., {Caraveo}, P.~A., {Pellizzoni}, A., {et~al.} 2007, \aap, 474,
  575

\bibitem[{{Frail} \& {Hjellming}(1991)}]{1991AJ....101.2126F}
{Frail}, D.~A. \& {Hjellming}, R.~M. 1991, \aj, 101, 2126

\bibitem[{{Gaia Collaboration} {et~al.}(2021){Gaia Collaboration}, {Brown},
  {Vallenari}, {Prusti}, {de Bruijne}, {Babusiaux}, {Biermann}, {Creevey},
  {Evans}, {Eyer}, {Hutton}, {Jansen}, {Jordi}, {Klioner}, {Lammers},
  {Lindegren}, {Luri}, {Mignard}, {Panem}, {Pourbaix}, {Randich}, {Sartoretti},
  {Soubiran}, {Walton}, {Arenou}, {Bailer-Jones}, {Bastian}, {Cropper},
  {Drimmel}, {Katz}, {Lattanzi}, {van Leeuwen}, {Bakker}, {Cacciari},
  {Casta{\~n}eda}, {De Angeli}, {Ducourant}, {Fabricius}, {Fouesneau},
  {Fr{\'e}mat}, {Guerra}, {Guerrier}, {Guiraud}, {Jean-Antoine Piccolo},
  {Masana}, {Messineo}, {Mowlavi}, {Nicolas}, {Nienartowicz}, {Pailler},
  {Panuzzo}, {Riclet}, {Roux}, {Seabroke}, {Sordo}, {Tanga}, {Th{\'e}venin},
  {Gracia-Abril}, {Portell}, {Teyssier}, {Altmann}, {Andrae}, {Bellas-Velidis},
  {Benson}, {Berthier}, {Blomme}, {Brugaletta}, {Burgess}, {Busso}, {Carry},
  {Cellino}, {Cheek}, {Clementini}, {Damerdji}, {Davidson}, {Delchambre},
  {Dell'Oro}, {Fern{\'a}ndez-Hern{\'a}ndez}, {Galluccio}, {Garc{\'\i}a-Lario},
  {Garcia-Reinaldos}, {Gonz{\'a}lez-N{\'u}{\~n}ez}, {Gosset}, {Haigron},
  {Halbwachs}, {Hambly}, {Harrison}, {Hatzidimitriou}, {Heiter},
  {Hern{\'a}ndez}, {Hestroffer}, {Hodgkin}, {Holl}, {Jan{\ss}en}, {Jevardat de
  Fombelle}, {Jordan}, {Krone-Martins}, {Lanzafame}, {L{\"o}ffler}, {Lorca},
  {Manteiga}, {Marchal}, {Marrese}, {Moitinho}, {Mora}, {Muinonen}, {Osborne},
  {Pancino}, {Pauwels}, {Petit}, {Recio-Blanco}, {Richards}, {Riello},
  {Rimoldini}, {Robin}, {Roegiers}, {Rybizki}, {Sarro}, {Siopis}, {Smith},
  {Sozzetti}, {Ulla}, {Utrilla}, {van Leeuwen}, {van Reeven}, {Abbas}, {Abreu
  Aramburu}, {Accart}, {Aerts}, {Aguado}, {Ajaj}, {Altavilla}, {{\'A}lvarez},
  {{\'A}lvarez Cid-Fuentes}, {Alves}, {Anderson}, {Anglada Varela}, {Antoja},
  {Audard}, {Baines}, {Baker}, {Balaguer-N{\'u}{\~n}ez}, {Balbinot}, {Balog},
  {Barache}, {Barbato}, {Barros}, {Barstow}, {Bartolom{\'e}}, {Bassilana},
  {Bauchet}, {Baudesson-Stella}, {Becciani}, {Bellazzini}, {Bernet}, {Bertone},
  {Bianchi}, {Blanco-Cuaresma}, {Boch}, {Bombrun}, {Bossini}, {Bouquillon},
  {Bragaglia}, {Bramante}, {Breedt}, {Bressan}, {Brouillet}, {Bucciarelli},
  {Burlacu}, {Busonero}, {Butkevich}, {Buzzi}, {Caffau}, {Cancelliere},
  {C{\'a}novas}, {Cantat-Gaudin}, {Carballo}, {Carlucci}, {Carnerero},
  {Carrasco}, {Casamiquela}, {Castellani}, {Castro-Ginard}, {Castro Sampol},
  {Chaoul}, {Charlot}, {Chemin}, {Chiavassa}, {Cioni}, {Comoretto}, {Cooper},
  {Cornez}, {Cowell}, {Crifo}, {Crosta}, {Crowley}, {Dafonte}, {Dapergolas},
  {David}, {David}, {de Laverny}, {De Luise}, {De March}, {De Ridder}, {de
  Souza}, {de Teodoro}, {de Torres}, {del Peloso}, {del Pozo}, {Delbo},
  {Delgado}, {Delgado}, {Delisle}, {Di Matteo}, {Diakite}, {Diener},
  {Distefano}, {Dolding}, {Eappachen}, {Edvardsson}, {Enke}, {Esquej}, {Fabre},
  {Fabrizio}, {Faigler}, {Fedorets}, {Fernique}, {Fienga}, {Figueras},
  {Fouron}, {Fragkoudi}, {Fraile}, {Franke}, {Gai}, {Garabato},
  {Garcia-Gutierrez}, {Garc{\'\i}a-Torres}, {Garofalo}, {Gavras}, {Gerlach},
  {Geyer}, {Giacobbe}, {Gilmore}, {Girona}, {Giuffrida}, {Gomel}, {Gomez},
  {Gonzalez-Santamaria}, {Gonz{\'a}lez-Vidal}, {Granvik},
  {Guti{\'e}rrez-S{\'a}nchez}, {Guy}, {Hauser}, {Haywood}, {Helmi}, {Hidalgo},
  {Hilger}, {H{\l}adczuk}, {Hobbs}, {Holland}, {Huckle}, {Jasniewicz},
  {Jonker}, {Juaristi Campillo}, {Julbe}, {Karbevska}, {Kervella}, {Khanna},
  {Kochoska}, {Kontizas}, {Kordopatis}, {Korn}, {Kostrzewa-Rutkowska},
  {Kruszy{\'n}ska}, {Lambert}, {Lanza}, {Lasne}, {Le Campion}, {Le Fustec},
  {Lebreton}, {Lebzelter}, {Leccia}, {Leclerc}, {Lecoeur-Taibi}, {Liao},
  {Licata}, {Lindstr{\o}m}, {Lister}, {Livanou}, {Lobel}, {Madrero Pardo},
  {Managau}, {Mann}, {Marchant}, {Marconi}, {Marcos Santos}, {Marinoni},
  {Marocco}, {Marshall}, {Martin Polo}, {Mart{\'\i}n-Fleitas}, {Masip},
  {Massari}, {Mastrobuono-Battisti}, {Mazeh}, {McMillan}, {Messina},
  {Michalik}, {Millar}, {Mints}, {Molina}, {Molinaro}, {Moln{\'a}r},
  {Montegriffo}, {Mor}, {Morbidelli}, {Morel}, {Morris}, {Mulone}, {Munoz},
  {Muraveva}, {Murphy}, {Musella}, {Noval}, {Ord{\'e}novic}, {Orr{\`u}},
  {Osinde}, {Pagani}, {Pagano}, {Palaversa}, {Palicio}, {Panahi}, {Pawlak},
  {Pe{\~n}alosa Esteller}, {Penttil{\"a}}, {Piersimoni}, {Pineau}, {Plachy},
  {Plum}, {Poggio}, {Poretti}, {Poujoulet}, {Pr{\v{s}}a}, {Pulone}, {Racero},
  {Ragaini}, {Rainer}, {Raiteri}, {Rambaux}, {Ramos}, {Ramos-Lerate}, {Re
  Fiorentin}, {Regibo}, {Reyl{\'e}}, {Ripepi}, {Riva}, {Rixon}, {Robichon},
  {Robin}, {Roelens}, {Rohrbasser}, {Romero-G{\'o}mez}, {Rowell}, {Royer},
  {Rybicki}, {Sadowski}, {Sagrist{\`a} Sell{\'e}s}, {Sahlmann}, {Salgado},
  {Salguero}, {Samaras}, {Sanchez Gimenez}, {Sanna}, {Santove{\~n}a},
  {Sarasso}, {Schultheis}, {Sciacca}, {Segol}, {Segovia}, {S{\'e}gransan},
  {Semeux}, {Shahaf}, {Siddiqui}, {Siebert}, {Siltala}, {Slezak}, {Smart},
  {Solano}, {Solitro}, {Souami}, {Souchay}, {Spagna}, {Spoto}, {Steele},
  {Steidelm{\"u}ller}, {Stephenson}, {S{\"u}veges}, {Szabados}, {Szegedi-Elek},
  {Taris}, {Tauran}, {Taylor}, {Teixeira}, {Thuillot}, {Tonello}, {Torra},
  {Torra}, {Turon}, {Unger}, {Vaillant}, {van Dillen}, {Vanel}, {Vecchiato},
  {Viala}, {Vicente}, {Voutsinas}, {Weiler}, {Wevers}, {Wyrzykowski}, {Yoldas},
  {Yvard}, {Zhao}, {Zorec}, {Zucker}, {Zurbach}, \&
  {Zwitter}}]{2021A&A...649A...1G}
{Gaia Collaboration}, {Brown}, A.~G.~A., {Vallenari}, A., {et~al.} 2021, \aap,
  649, A1

\bibitem[{{Goldoni} \& {Mereghetti}(1995)}]{1995A&A...299..751G}
{Goldoni}, P. \& {Mereghetti}, S. 1995, \aap, 299, 751

\bibitem[{{Gregory}(2002)}]{2002ApJ...575..427G}
{Gregory}, P.~C. 2002, \apj, 575, 427

\bibitem[{{Gregory} \& {Taylor}(1978)}]{1978Natur.272..704G}
{Gregory}, P.~C. \& {Taylor}, A.~R. 1978, \nat, 272, 704

\bibitem[{{Grundstrom} {et~al.}(2007){Grundstrom}, {Caballero-Nieves}, {Gies},
  {Huang}, {McSwain}, {Rafter}, {Riddle}, {Williams}, \&
  {Wingert}}]{2007ApJ...656..437G}
{Grundstrom}, E.~D., {Caballero-Nieves}, S.~M., {Gies}, D.~R., {et~al.} 2007,
  \apj, 656, 437

\bibitem[{Gupta \& Boettcher(2006)}]{Gupta2006ATL}
Gupta, S. \& Boettcher, M. 2006, The Astrophysical Journal, 650

\bibitem[{{Hadasch} {et~al.}(2012){Hadasch}, {Torres}, {Tanaka}, {Corbet},
  {Hill}, {Dubois}, {Dubus}, {Glanzman}, {Corbel}, {Li}, {Chen}, {Zhang},
  {Caliandro}, {Kerr}, {Richards}, {Max-Moerbeck}, {Readhead}, \&
  {Pooley}}]{2012ApJ...749...54H}
{Hadasch}, D., {Torres}, D.~F., {Tanaka}, T., {et~al.} 2012, \apj, 749, 54

\bibitem[{{Harrison} {et~al.}(2000){Harrison}, {Ray}, {Leahy}, {Waltman}, \&
  {Pooley}}]{2000ApJ...528..454H}
{Harrison}, F.~A., {Ray}, P.~S., {Leahy}, D.~A., {Waltman}, E.~B., \& {Pooley},
  G.~G. 2000, \apj, 528, 454

\bibitem[{{Hutchings} \& {Crampton}(1981)}]{1981PASP...93..486H}
{Hutchings}, J.~B. \& {Crampton}, D. 1981, \pasp, 93, 486

\bibitem[{Leahy {et~al.}(1997)Leahy, Harrison, \& Yoshida}]{Leahy1997TheAX}
Leahy, D.~A., Harrison, F.~A., \& Yoshida, A. 1997, The Astrophysical Journal,
  475, 823

\bibitem[{{Li} {et~al.}(2018){Li}, {Torres}, {Lin}, {Grondin}, {Kerr},
  {Lemoine-Goumard}, \& {de O{\~n}a Wilhelmi}}]{2018Jian}
{Li}, J., {Torres}, D.~F., {Lin}, T.~T., {et~al.} 2018, \apj, 858, 84

\bibitem[{{Li} {et~al.}(2011){Li}, {Torres}, {Zhang}, {Chen}, {Hadasch}, {Ray},
  {Kretschmar}, {Rea}, \& {Wang}}]{2011ApJ...733...89L}
{Li}, J., {Torres}, D.~F., {Zhang}, S., {et~al.} 2011, \apj, 733, 89

\bibitem[{{Li} {et~al.}(2012){Li}, {Torres}, {Zhang}, {Hadasch}, {Rea},
  {Caliandro}, {Chen}, \& {Wang}}]{2012ApJ...744L..13L}
{Li}, J., {Torres}, D.~F., {Zhang}, S., {et~al.} 2012, \apjl, 744, L13

\bibitem[{{Lightkurve Collaboration} {et~al.}(2018){Lightkurve Collaboration},
  {Cardoso}, {Hedges}, {Gully-Santiago}, {Saunders}, {Cody}, {Barclay}, {Hall},
  {Sagear}, {Turtelboom}, {Zhang}, {Tzanidakis}, {Mighell}, {Coughlin}, {Bell},
  {Berta-Thompson}, {Williams}, {Dotson}, \& {Barentsen}}]{2018ascl.soft12013L}
{Lightkurve Collaboration}, {Cardoso}, J.~V.~d.~M., {Hedges}, C., {et~al.}
  2018, {Lightkurve: Kepler and TESS time series analysis in Python},
  Astrophysics Source Code Library

\bibitem[{{Maraschi} \& {Treves}(1981)}]{1981MNRAS.194P...1M}
{Maraschi}, L. \& {Treves}, A. 1981, \mnras, 194, 1P

\bibitem[{{Mart{\'\i}} {et~al.}(2017){Mart{\'\i}}, {Luque-Escamilla}, \&
  {Garc{\'\i}a-Hern{\'a}ndez}}]{2017BlgAJ..26...91M}
{Mart{\'\i}}, J., {Luque-Escamilla}, P.~L., \& {Garc{\'\i}a-Hern{\'a}ndez},
  M.~T. 2017, Bulgarian Astronomical Journal, 26, 91

\bibitem[{{Mart\'{\i}} \& {Paredes}(1995)}]{1995A&A...298..151M}
{Mart\'{\i}}, J. \& {Paredes}, J.~M. 1995, \aap, 298, 151

\bibitem[{{Massi} {et~al.}(2017){Massi}, {Migliari}, \&
  {Chernyakova}}]{2017MNRAS.468.3689M}
{Massi}, M., {Migliari}, S., \& {Chernyakova}, M. 2017, \mnras, 468, 3689

\bibitem[{{Massi} {et~al.}(2004){Massi}, {Rib{\'o}}, {Paredes}, {Garrington},
  {Peracaula}, \& {Mart{\'\i}}}]{2004A&A...414L...1M}
{Massi}, M., {Rib{\'o}}, M., {Paredes}, J.~M., {et~al.} 2004, \aap, 414, L1

\bibitem[{{Massi} {et~al.}(2001){Massi}, {Rib{\'o}}, {Paredes}, {Peracaula}, \&
  {Estalella}}]{2001A&A...376..217M}
{Massi}, M., {Rib{\'o}}, M., {Paredes}, J.~M., {Peracaula}, M., \& {Estalella},
  R. 2001, \aap, 376, 217

\bibitem[{{Mattox} {et~al.}(1996){Mattox}, {Bertsch}, {Chiang}, {Dingus},
  {Digel}, {Esposito}, {Fierro}, {Hartman}, {Hunter}, {Kanbach}, {Kniffen},
  {Lin}, {Macomb}, {Mayer-Hasselwander}, {Michelson}, {von Montigny},
  {Mukherjee}, {Nolan}, {Ramanamurthy}, {Schneid}, {Sreekumar}, {Thompson}, \&
  {Willis}}]{1996ApJ...461..396M}
{Mattox}, J.~R., {Bertsch}, D.~L., {Chiang}, J., {et~al.} 1996, \apj, 461, 396

\bibitem[{{Mendelson} \& {Mazeh}(1989)}]{1989MNRAS.239..733M}
{Mendelson}, H. \& {Mazeh}, T. 1989, \mnras, 239, 733

\bibitem[{{Metzger} {et~al.}(2015){Metzger}, {Finzell}, {Vurm}, {Hasco{\"e}t},
  {Beloborodov}, \& {Chomiuk}}]{2015MNRAS.450.2739M}
{Metzger}, B.~D., {Finzell}, T., {Vurm}, I., {et~al.} 2015, \mnras, 450, 2739

\bibitem[{{Metzger} {et~al.}(2014){Metzger}, {Hasco{\"e}t}, {Vurm},
  {Beloborodov}, {Chomiuk}, {Sokoloski}, \& {Nelson}}]{2014MNRAS.442..713M}
{Metzger}, B.~D., {Hasco{\"e}t}, R., {Vurm}, I., {et~al.} 2014, \mnras, 442,
  713

\bibitem[{{Paredes}(1987)}]{1987PhDT.......113P}
{Paredes}, J.~M. 1987, PhD thesis, University of Barcelona

\bibitem[{{Paredes} \& {Figueras}(1986)}]{1986A&A...154L..30P}
{Paredes}, J.~M. \& {Figueras}, F. 1986, \aap, 154, L30

\bibitem[{{Paredes} {et~al.}(1997){Paredes}, {Marti}, {Peracaula}, \&
  {Ribo}}]{1997A&A...320L..25P}
{Paredes}, J.~M., {Marti}, J., {Peracaula}, M., \& {Ribo}, M. 1997, \aap, 320,
  L25

\bibitem[{{Paredes} {et~al.}(2007){Paredes}, {Rib{\'o}}, {Bosch-Ramon}, {West},
  {Butt}, {Torres}, \& {Mart{\'\i}}}]{2007ApJ...664L..39P}
{Paredes}, J.~M., {Rib{\'o}}, M., {Bosch-Ramon}, V., {et~al.} 2007, \apjl, 664,
  L39

\bibitem[{{Peracaula} {et~al.}(1997){Peracaula}, {Marti}, \&
  {Paredes}}]{1997A&A...328..283P}
{Peracaula}, M., {Marti}, J., \& {Paredes}, J.~M. 1997, \aap, 328, 283

\bibitem[{{Reig} \& {Fabregat}(2015)}]{2015A&A...574A..33R}
{Reig}, P. \& {Fabregat}, J. 2015, \aap, 574, A33

\bibitem[{Romero {et~al.}(2005)Romero, Christiansen, \&
  Orellana}]{Romero2005HadronicHG}
Romero, G.~E., Christiansen, H.~R., \& Orellana, M. 2005, The Astrophysical
  Journal, 632, 1093

\bibitem[{Romero {et~al.}(2007)Romero, Okazaki, Orellana, \&
  Owocki}]{Romero2007AccretionVC}
Romero, G.~E., Okazaki, A.~T., Orellana, M., \& Owocki, S.~P. 2007, Astronomy
  and Astrophysics, 474, 15

\bibitem[{{Sidoli} {et~al.}(2006){Sidoli}, {Pellizzoni}, {Vercellone},
  {Moroni}, {Mereghetti}, \& {Tavani}}]{2006A&A...459..901S}
{Sidoli}, L., {Pellizzoni}, A., {Vercellone}, S., {et~al.} 2006, \aap, 459, 901

\bibitem[{{Steinberg} \& {Metzger}(2018)}]{2018MNRAS.479..687S}
{Steinberg}, E. \& {Metzger}, B.~D. 2018, \mnras, 479, 687

\bibitem[{{Taylor} \& {Gregory}(1982)}]{1982ApJ...255..210T}
{Taylor}, A.~R. \& {Gregory}, P.~C. 1982, \apj, 255, 210

\bibitem[{{Taylor} {et~al.}(1996){Taylor}, {Young}, {Peracaula}, {Kenny}, \&
  {Gregory}}]{1996A&A...305..817T}
{Taylor}, A.~R., {Young}, G., {Peracaula}, M., {Kenny}, H.~T., \& {Gregory},
  P.~C. 1996, \aap, 305, 817

\bibitem[{Torres \& Halzen(2007)}]{torres2007ls}
Torres, D.~F. \& Halzen, F. 2007, Astroparticle Physics, 27, 500

\bibitem[{{Torres} {et~al.}(2012){Torres}, {Rea}, {Esposito}, {Li}, {Chen}, \&
  {Zhang}}]{2012ApJ...744..106T}
{Torres}, D.~F., {Rea}, N., {Esposito}, P., {et~al.} 2012, \apj, 744, 106

\bibitem[{{Waters} {et~al.}(1988){Waters}, {van den Heuvel}, {Taylor},
  {Habets}, \& {Persi}}]{1988A&A...198..200W}
{Waters}, L.~B.~F.~M., {van den Heuvel}, E.~P.~J., {Taylor}, A.~R., {Habets},
  G.~M.~H.~J., \& {Persi}, P. 1988, \aap, 198, 200

\bibitem[{{Weng} {et~al.}(2022){Weng}, {Qian}, {Wang}, {Torres}, {Papitto},
  {Jiang}, {Xu}, {Li}, {Yan}, {Liu}, {Ge}, \& {Yuan}}]{2022NatAs.tmp...71W}
{Weng}, S.-S., {Qian}, L., {Wang}, B.-J., {et~al.} 2022, Nature Astronomy
  [\eprint[arXiv]{2203.09423}]

\bibitem[{{Wood} {et~al.}(2017){Wood}, {Caputo}, {Charles}, {Di Mauro},
  {Magill}, {Perkins}, \& {Fermi-LAT Collaboration}}]{2017ICRC...35..824W}
{Wood}, M., {Caputo}, R., {Charles}, E., {et~al.} 2017, in International Cosmic
  Ray Conference, Vol. 301, 35th International Cosmic Ray Conference
  (ICRC2017), 824

\bibitem[{{Zamanov} {et~al.}(2014){Zamanov}, {Mart{\'\i}}, {Stoyanov},
  {Borissova}, \& {Tomov}}]{2014A&A...561L...2Z}
{Zamanov}, R., {Mart{\'\i}}, J., {Stoyanov}, K., {Borissova}, A., \& {Tomov},
  N.~A. 2014, \aap, 561, L2

\bibitem[{Zdziarski {et~al.}(2010)Zdziarski, Neronov, \&
  Chernyakova}]{2010MNRAS.403.1873Z}
Zdziarski, A.~A., Neronov, A., \& Chernyakova, M. 2010, Mon. Not. Roy. Astron.
  Soc., 403, 1873

\end{thebibliography}

%\begin{thebibliography}{}

  %\bibitem[\protect\citeauthoryear{Torres et al.}{2017}]{2017ApJ...836...68T} Torres D.~F., Ji L., Li J., Papitto A., Rea N., de O{\~n}a Wilhelmi E., Zhang S., 2017, ApJ, 836, 68. doi:10.3847/1538-4357/836/1/68
  %\bibitem[2021]{paper2021} paper, 2021.

 %\end{thebibliography}

\end{document}